# The impact of Monte Carlo simulation: a scientometric analysis of scholarly literature

Maria Grazia PIA[1*], Tullio BASAGLIA[2], Zane W. BELL[3], Paul V. DRESSENDORFER[4]

[1]*INFN Sezione di Genova, 16146 Genova, Italy*
[2]*CERN, 1211 Geneva, Switzerland*
[3]*ORNL, Oak Ridge, TN 37830, USA*
[4]*IEEE, Piscataway, NJ 08854, USA*

A scientometric analysis of Monte Carlo simulation and Monte Carlo codes has been performed over a set of representative scholarly journals related to radiation physics. The results of this study are reported and discussed. They document and quantitatively appraise the role of Monte Carlo methods and codes in scientific research and engineering applications.

***KEYWORDS: Monte Carlo, EGS, FLUKA, GEANT, Geant4, MCNP, Penelope***

## I. Introduction

The use of Monte Carlo methods to simulate particle interactions with matter has increased significantly in the recent years, and nowadays Monte Carlo simulation is an essential research tool in such diverse fields as nuclear and particle physics, astrophysics and space science, medical physics, radiation protection, electronic components development etc.

Large-scale Monte Carlo codes, like MCNP[1)2)3)], GEANT[4)], Geant4[5)6)] and EGS[7)8)9)], are widely present in scholarly literature, while citations to several other codes, often addressing specific application domains, document the important role of Monte Carlo simulation in many areas of physics and engineering literature.

A scientometric analysis has been performed over a set of scholarly journals in various fields related to radiation physics. The study spans five decades and concerns both fundamental physics and technological applications. The results highlight the evolution of technological research in nuclear and particle physics, and related fields, and the impact of Monte Carlo simulation software in the experimental realm.

## II. Monte Carlo in scholarly journals

The analysis involved a number of representative journals in instrumentation, fundamental physics, astrophysics and medical physics: IEEE Transactions on Nuclear Science (TNS), Nuclear Instruments and Methods (NIM) A and B, Nuclear Physics B, Physics Letters B, Physical Review D, Physical Review Letters, Medical Physics, Physics in Medicine and Biology, and the Astrophysical Journal. The total number of papers published in these journals in the period covered by the scientometric analysis is shown in **Figure 1**.

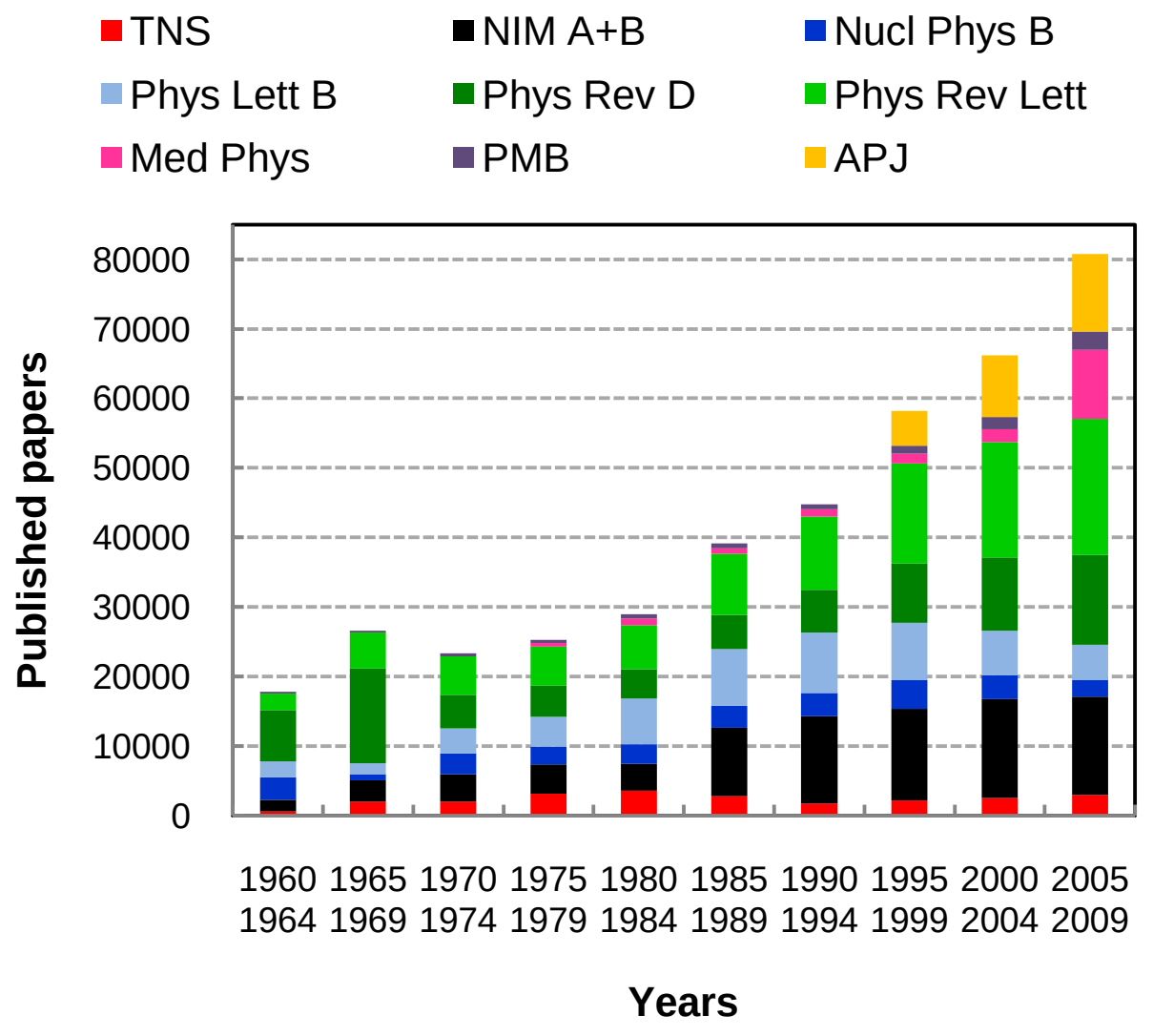

**Fig. 1** Number of papers published between 1960 and 2009 by the journals considered in the scientometric analysis.

The analysis was performed over five decades (1960-2009). Some journals provide full-text online search facilities over shorter periods, or the authors' library subscriptions did not cover the whole range of fifty years; for them the analysis was necessarily limited to the accessible time interval. The journal with limited time coverage can be easily identified in the following plots.

*Corresponding Author, E-mail:MariaGrazia.Pia@ge.infn.it

Two searches were performed: the occurrence of the "Monte Carlo" string, and the occurrence of either "Monte Carlo" or "simulation" in the text of published articles. The results are summarized in the following figures.

An evident trend of increasing number of occurrences of the "Monte Carlo" string in the analyzed sample of journals is evident in **Figure 2**. However, the total number of articles published in the selected journals has also increased.

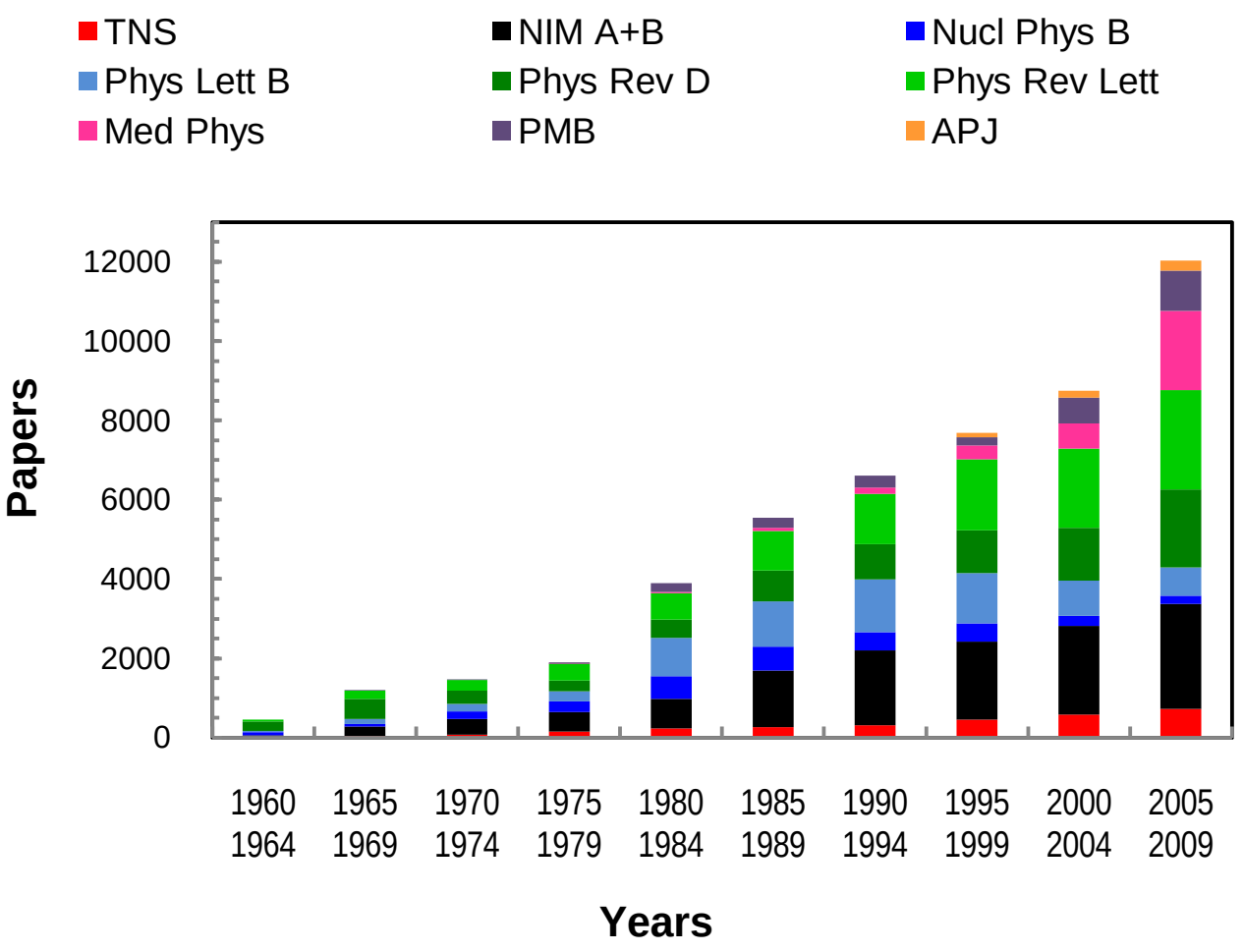

**Fig. 2** Absolute number of papers mentioning "Monte Carlo" in representative scholarly journals; each bin groups five years between 1960 and 2009.

**Figures 3-5** show the percentage of papers mentioning "Monte Carlo" respectively in representative instrumentation, fundamental physics and medical physics journals.

A trend towards increasing fraction of papers mentioning "Monte Carlo" is visible in instrumentation journals and in multi-disciplinary fundamental physics journals (Physics Letters and Physical Review Letters), while the trend is more controversial in particle physics journals (Physical Review D and Nuclear Physics B) and in medical physics journals. It should be noted that the Nuclear Physics B sample includes papers published in Nuclear Physics before the scope of this journal was split between Nuclear Physics A and B in 1968.

The fraction of papers mentioning "Monte Carlo" increases from a few percent in the 60's to approximately 15-25% in instrumentation journals, and to approximately 12% in fundamental physics journals; it is of the order of a few percent in the analyzed astrophysics journal. The presence of "Monte Carlo" was negligible in medical physics journals until the late 70's, but it has increased to more than 30% of the total number of papers in the following years.

The distribution of papers mentioning "Monte Carlo" in the sample of analyzed journals is illustrated in **Figure 6**; it is normalized to the total number of papers which mention this string. It appears that papers associated with fundamental physics research carry the largest weight in the sample, followed by papers published in instrumentation journals and in medical physics.

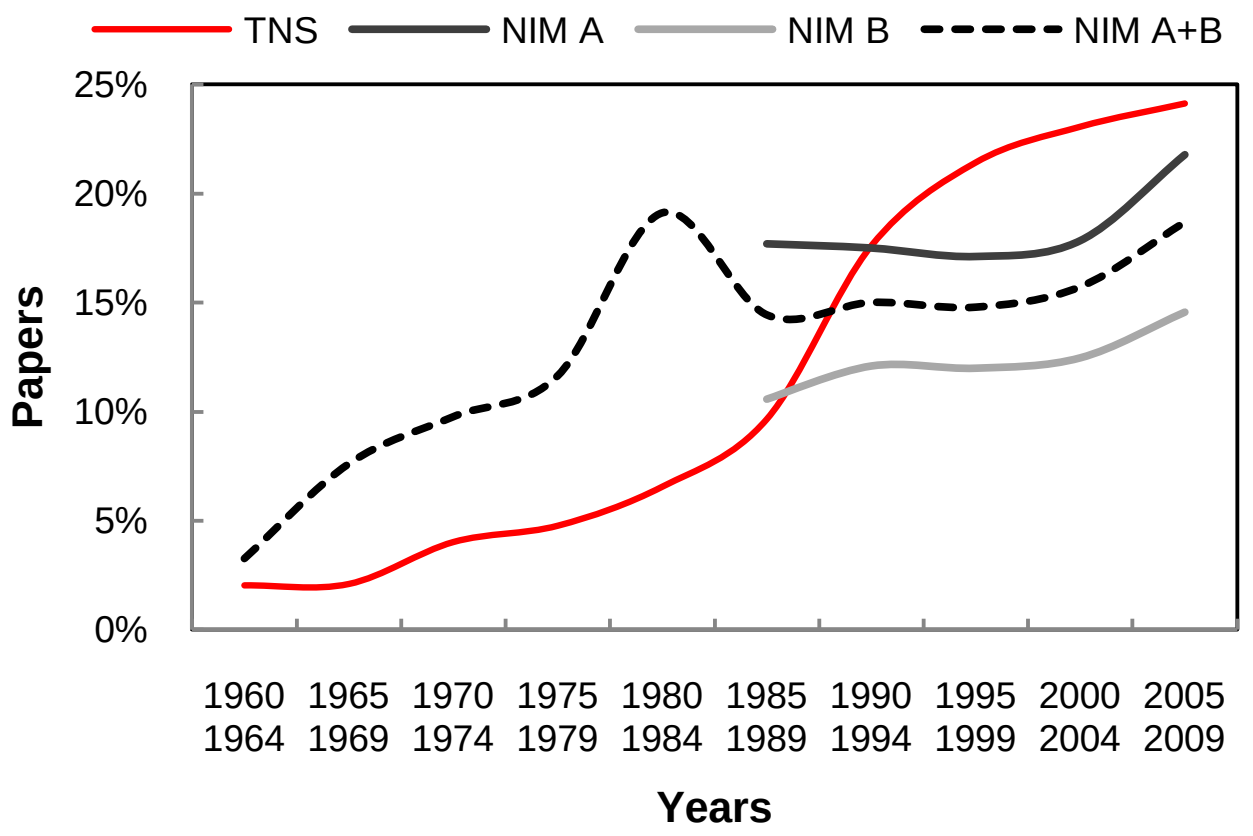

**Fig. 3** Percentage of papers mentioning "Monte Carlo" in instrumentation journals with respect to the total number of papers published in the five year period corresponding to each bin.

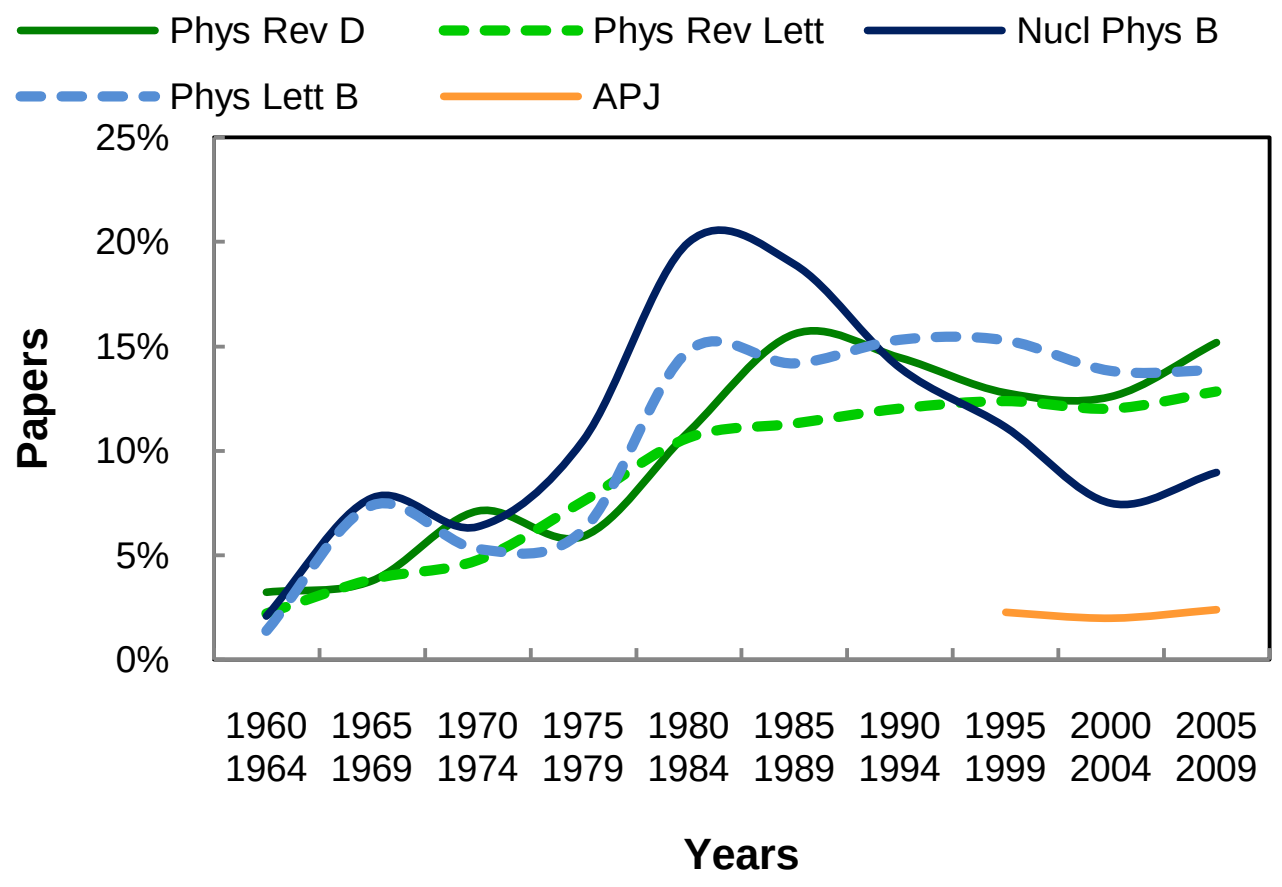

**Fig. 4** Percentage of papers mentioning "Monte Carlo" in fundamental physics journals with respect to the total number of papers published in the five year period corresponding to each bin.

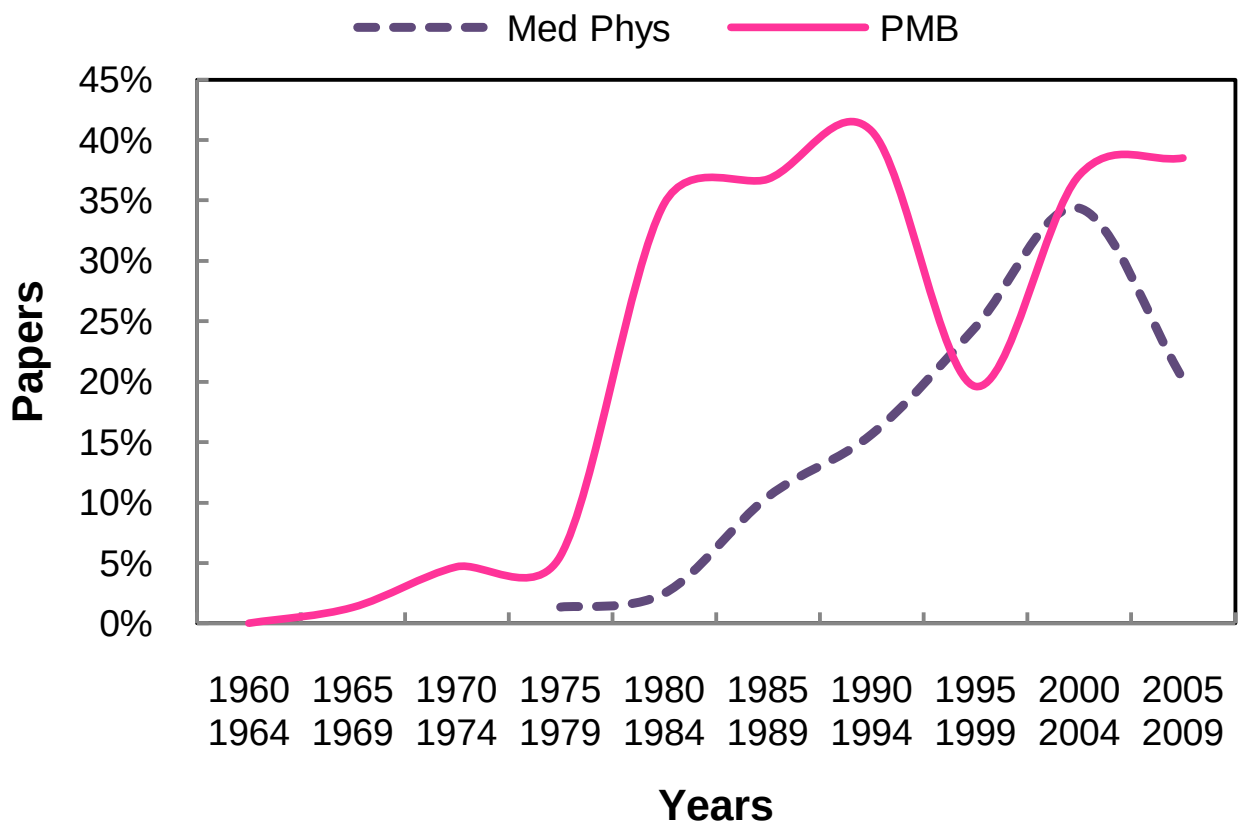

**Fig. 5** Percentage of papers mentioning "Monte Carlo" in medical physics journals with respect to the total number of papers published in the five year period corresponding to each bin.

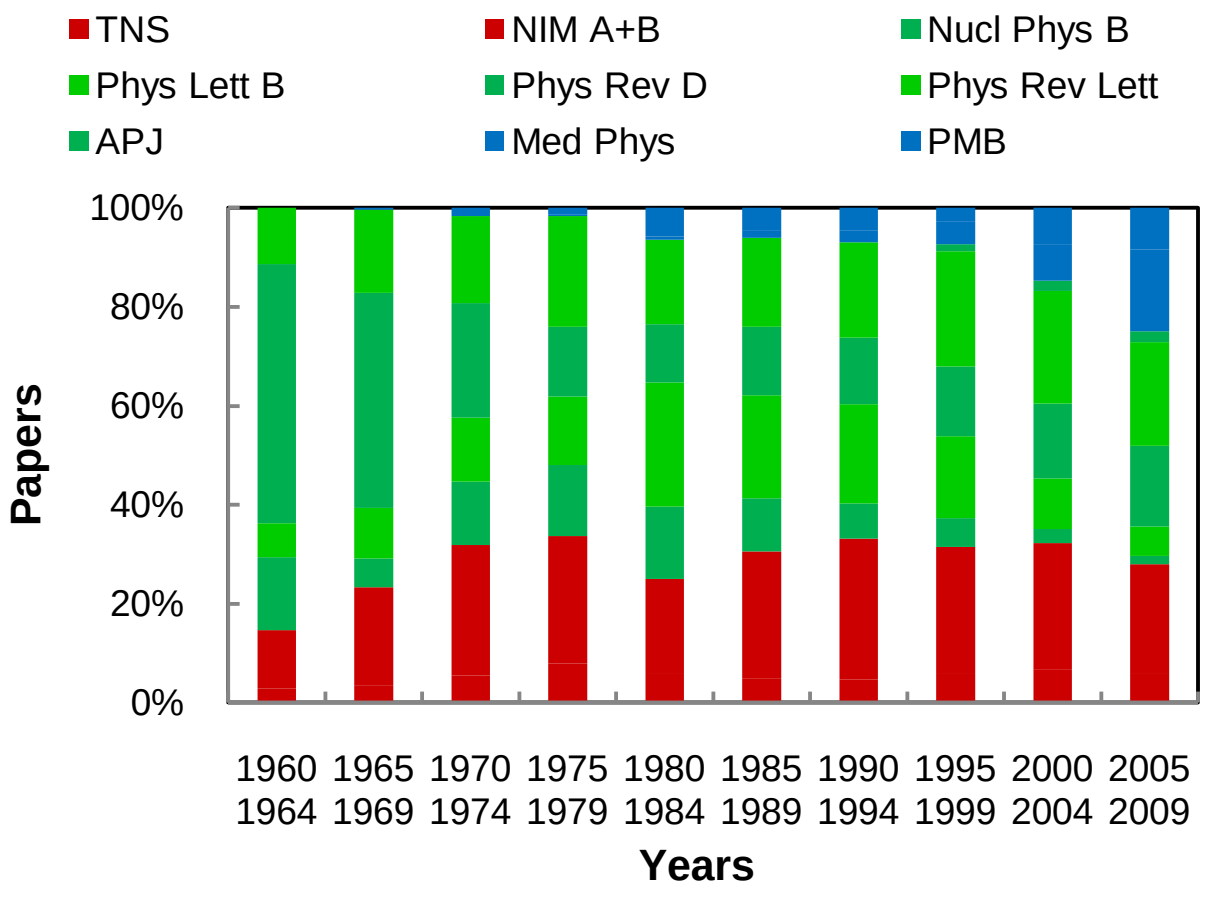

**Fig. 6** The distribution of papers mentioning "Monte Carlo" in the sample of analyzed journals, normalized to the total number of papers where the string occurs: instrumentation journals (dark red), fundamental physics and astrophysics journals (green) and medical physics journals (blue).

The analysis of the occurrence of either "Monte Carlo" or "simulation" in the text of published papers was practically possible in a subset of journals, whose web site interfaces support Boolean searches. It shows similar trends to the previous analysis, although the fraction of papers selected through this extended search pattern is larger.

An example is shown in **Figure 7**, which concerns instrumentation journals. One can observe that in recent years approximately half of the papers published in TNS, and 40% of those published in NIM, mention Monte Carlo or simulation; the growth of the relevance of simulation in instrumentation research is clearly visible, when the results are compared over fifty years.

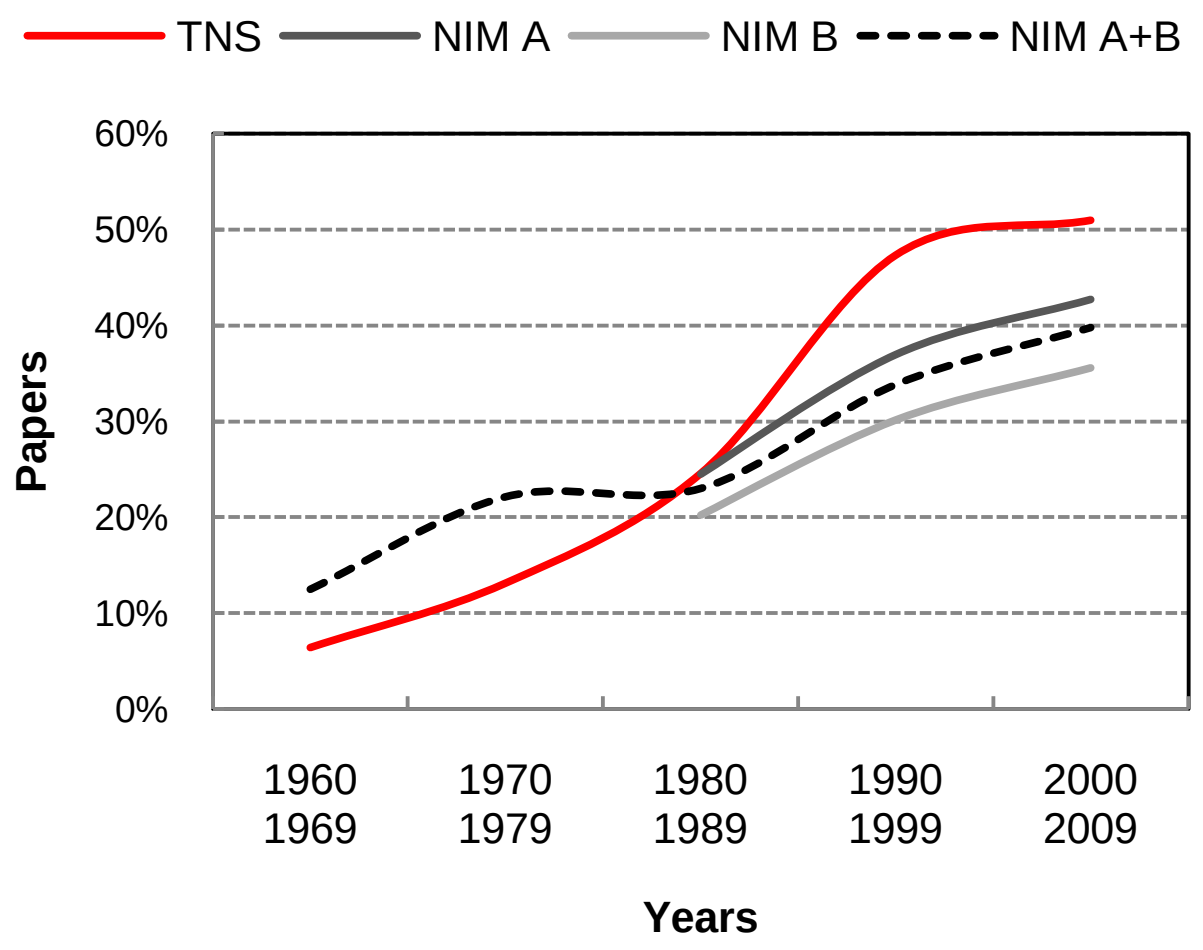

**Fig. 7** Percentage of papers mentioning "Monte Carlo" or "simulation" in instrumentation journals with respect to the total number of papers published in the ten year period corresponding to each bin.

Fundamental physics journals exhibit a similar trend of increasing number of papers mentioning "Monte Carlo" or "simulation", with the exception of Nuclear Physics B; the data are plotted in **Figure 8**.

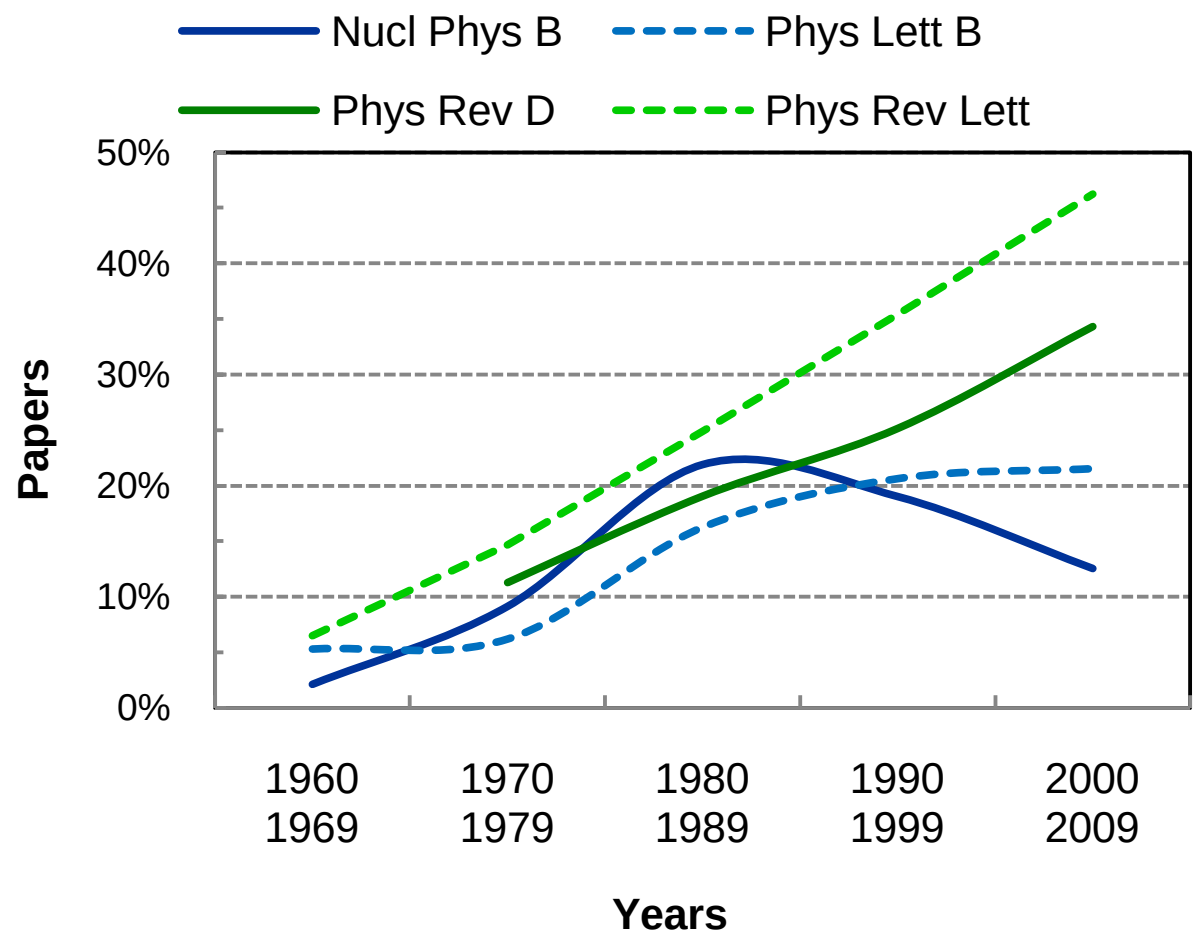

**Fig. 8** Percentage of papers mentioning "Monte Carlo" or "simulation" in fundamental physics journals with respect to the total number of papers published in the ten year period corresponding to each bin.

## III. Monte Carlo codes

A set of well known Monte Carlo codes was considered in this analysis: EGS[7)8)9)], FLUKA[10)11)], GEANT[4)] and Geant4[5)6)], MCNP[1)2)3)] and Penelope[12)]. This selection is representative of the field; it is not intended to be exhaustive.

Most of these codes cannot be associated with a reference publication in a journal; therefore, their role in scholarly literature cannot be appraised through the analysis of pertinent citations. The evaluation of their presence in the literature reported in this paper was based on the mention of the codes in the literature; this analysis was necessarily limited to publishers providing full-text search tools through their web interfaces. The analysis concerned the same journals listed in the previous section over the same time frame; in this context it should be taken into account that some Monte Carlo codes became publicly available only in recent years (e.g. Geant4 was first released in December 1998).

The data collection looked for the occurrence of strings associated with the name of the various codes in published articles; the search pattern took into account different versions of the codes and naming variants (e.g. Geant4 and Geant-4). EGS and MCNP group data associated with different branches of these codes (e.g. MCNP and MCNPX).

It should be noted that the collected data samples contain some mismatched GEANT attributions, since in some publications Geant4 is erroneously identified as GEANT. Similarly, in some cases FLUKA refers to an early version of that code interfaced to GEANT as a hadronic physics package, rather than to the standalone code. Whenever possible these mismatched attributions were corrected in the statistical analysis, but, due to the limited online search interfaces provided by the publishers, their complete correction would require manually verifying the content of all the collected papers, which is obviously impractical.

The results are shown in **Figures 9** and **10**, which concern the last two decades. The number of times the selected Monte Carlo codes are mentioned before 1990 is negligible.

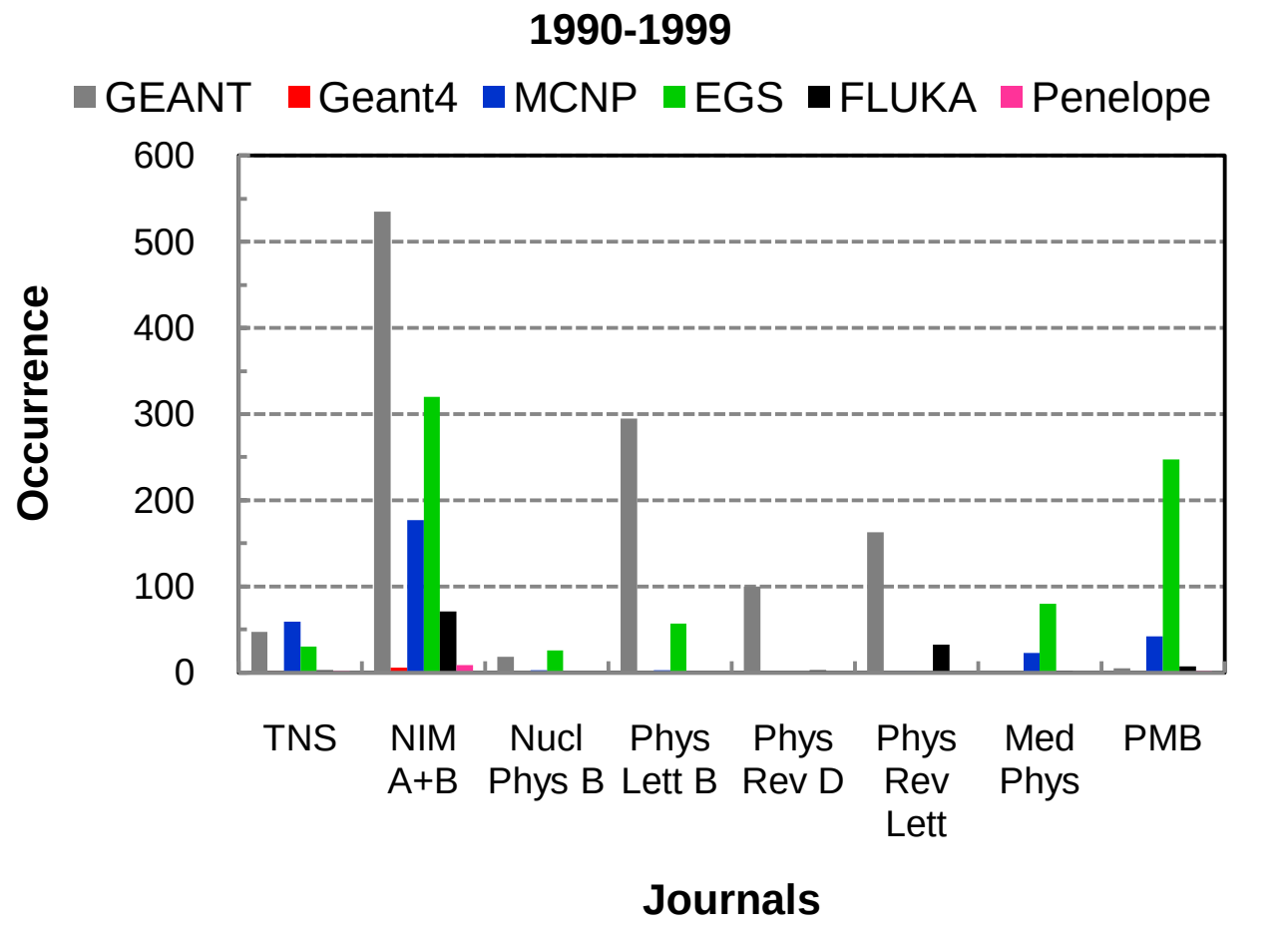

**Fig. 9** Absolute number of times selected large scale Monte Carlo codes are mentioned in representative journals in the 1990-1999 decade.

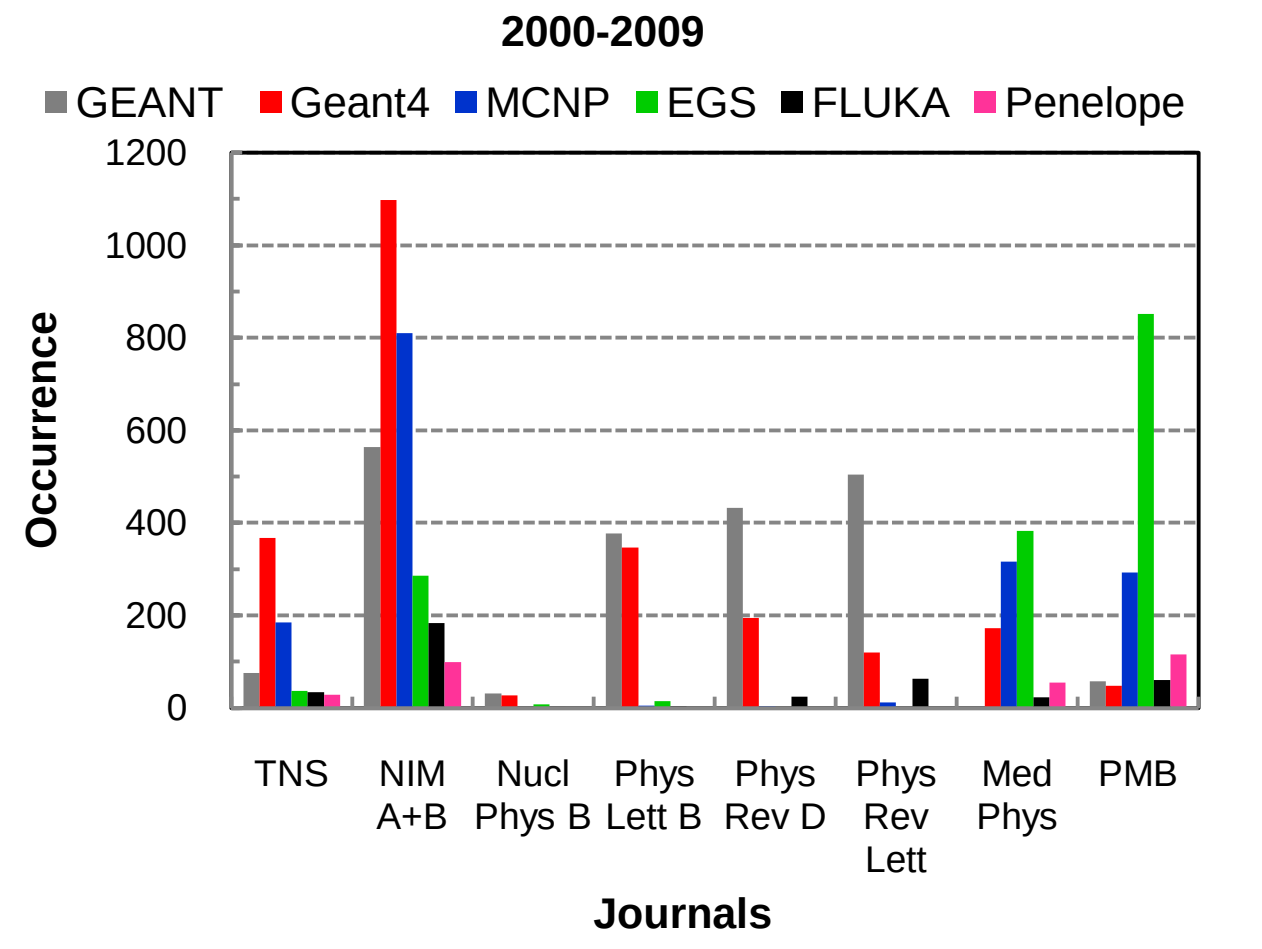

**Fig. 10** Absolute number of times selected large scale Monte Carlo codes are mentioned in representative journals in the 2000-2009 decade.

Among the selected Monte Carlo systems, in the 90's GEANT appeared to be the most widely mentioned code in fundamental physics journals and in NIM, while EGS was the most often mentioned one in medical physics journals.

In the last decade MCNP has increased its relative weight in instrumentation and medical physics journals. Geant4 appears to be the most frequently mentioned code in instrumentation journals, and EGS the most often mentioned in medical physics journals. GEANT is still frequently mentioned in fundamental physics journals, since various particle and nuclear physics experiments that started taking data in the previous years (e.g. experiments at the Tevatron and at LEP) did not upgrade their simulation configuration to more modern codes or versions in later publications to avoid introducing possible systematic effects in their physics results.

## IV. The record Monte Carlo paper

The Geant4 reference article published in 2003 has become the most cited paper in the whole Nuclear Science and Technology category of Thomson-Reuter's Journal Citation Reports[13], the official reference for impact factor determination; at the time of writing this paper (August 2010) it has crossed the threshold of 2000 citations. Thomson-Reuter classified it among the "current classic" paper selections.

This paper is currently the second most cited[14] article among the publications authored by two major research institutes, CERN and INFN, in the past two decades.

This outstanding performance contrasts with the tradition of nuclear science and technology research, which is largely dominated by hardware – rather than software – R&D (research and development), and with the relatively lower representation of software articles in scholarly literature[15] concerning particle and nuclear physics, and related disciplines.

A detailed analysis of some relevant features associated with this publication can be found in a recent paper[16].

## III. Conclusion

The scientometric analysis presented in the previous sections shows that Monte Carlo simulation of particle transport in matter plays a major role in scientific research. The presence of Monte Carlo methods has significantly increased in all the fields associated with the representative journals considered in this study.

All the large scale Monte Carlo systems evaluated in this paper exhibit a similar trend of increasing presence in scholarly literature. Some codes are especially relevant in specific publication domains (e.g. EGS in medical physics journals); others support multi-disciplinary applications documented in a variety of publication domains.